# Incentivizing the Workers for Truth Discovery in Crowdsourcing with Copiers

Lingyun Jiang[1], Xiaofu Niu[1], Jia Xu[1], Dejun Yang[1,2], Lijie Xu[1]

[1]Jiangsu Key Laboratory of Big Data Security and Intelligent Processing, Nanjing University of Posts and Telecommunications, Nanjing, Jiangsu 210023 China

[2]Department of Computer Science, Colorado School of Mines, Golden, CO 80401 USA

Email: jianglingyun@njupt.edu.cn, nxf75@foxmail.com, xujia@njupt.edu.cn, djyang@mines.edu, ljxu@njupt.edu.cn

Corresponding author: Jia Xu

*Abstract*—Crowdsourcing has become an efficient paradigm for performing large scale tasks. Truth discovery and incentive mechanism are fundamentally important for the crowdsourcing system. Many truth discovery methods and incentive mechanisms for crowdsourcing have been proposed. However, most of them cannot be applied to deal with the crowdsourcing with copiers. To address the issue, we formulate the problem of maximizing the social welfare such that all tasks can be completed with the least confidence for truth discovery. We design an incentive mechanism consisting truth discovery stage and reverse auction stage. In truth discovery stage, we estimate the truth for each task based on both the dependence and accuracy of workers. In reverse auction stage, we design a greedy algorithm to select the winners and determine the payment. Through both rigorous theoretical analysis and extensive simulations, we demonstrate that the proposed mechanisms achieve computational efficiency, individual rationality, truthfulness, and guaranteed approximation. Moreover, our truth discovery method shows prominent advantage in terms of precision when there are copiers in the crowdsourcing systems.

*Keywords—crowdsourcing; truth discovery; incentive mechanism; reverse auction; Bayesian analysis*

## I. INTRODUCTION

Crowdsourcing is a distributed problem-solving model, in which a crowd of undefined size is engaged to solve the complex problems through an open platform. Wikipedia [1], Zhihu [2], Freebase [3], and other knowledge repositories were created by workers, who contributed knowledge on a wide variety of topics. In recent years, crowdsourcing has been widely used in many fields, including video analysis [4], knowledge discovery [5], and Smart Citizen [6], conducting human-robot interaction studies [7]. With the rapid proliferation of smartphones integrated with a variety of embedded sensors, mobile crowdsourcing has become an efficient approach of data acquisition in large-scale sensing applications, such as photo selection [8], public bike trip selection [9], and indoor positioning systems [10].

Many crowdsourcing applications require integrating data from multiple workers, each of which provides a set of values as "facts". However, "facts and truth really don't have much to do with each other" [11]. Different workers may provide conflicting values, some being true while some being false. To provide data with high accuracy to the requesters, it is critical for the truth discovery systems to resolve conflicts and discover true values.

The crowdsourcer aggregates and extracts crowdsourced information in order to discover the truth, where the accuracy of crowdsourcing data is fundamentally important. In crowdsourcing, the accuracy of data can be largely affected by the expertise and willingness of individual workers [12, 13]. Particularly, the workers with different spatial-temporal contexts and personal effort levels usually submit data with different accuracy. Furthermore, the rational workers tend to strategically minimize their efforts when performing the tasks, and thus may degrade the accuracy of data.

Typically, we often expect the true value provided by more workers than any particular false one, so we can apply voting [14] and take the value provided by the majority of the workers as the truth. The main drawback of this approach is that they treat the reliability of each worker equally. Unfortunately, the behavior of copying between workers is common in practice [15], especially when the crowdsourcing tasks are in the form of questionnaire.

In a variety of domains, such as science, business, politics, arts, there are a huge number of workers to provide information, and a large part of the provided information exists repetition. Most of the information is about some static aspects of the world, such as the authors and publishers of books, directors, the actors and actresses of movies, and the presidents of a company in past years. In this scenario, the workers may copy, crawl, or aggregate data from other workers, and submit the copied data without the declaration of ownership.

TABLE 1: AN EXAMPLE OF CONFLICTING VALUES PROVIDED BY CROWDSOURCING WORKERS WITH COPIERS

| Workers / Tasks | 1 | 2 | 3 | 4 | 5 |
|---|---|---|---|---|---|
| *Stonebraker* | MIT | Berkeley | MIT | MIT | MS |
| *Dewitt* | MSR | MSR | UWise | UWisc | UWisc |
| *Bernstein* | MSR | MSR | MSR | MSR | MSR |
| *Carey* | UCI | AT&T | BEA | BEA | BEA |
| *Halevy* | Google | Google | UW | UW | UW |

The existence of copiers would invalidate the most of the existing truth discovery methods [31, 32, 35-37], since they consider that workers are independent of each other. For example, as shown in Tab. 1, there are five workers, who provide the affiliations of five researchers, and only worker 1 provides all correct data. However, since the affiliations provided by worker 4 and worker 5 are copied from worker 3 (with certain errors during copying), the naive voting method would consider them as the majority, making wrong decisions of the truth for *Dewitt*, *Carey*, and *Halevy*.

In this paper, we aim to develop an integrated solution to solve the following two issues: given the conflicting values provided by crowdsourcing workers with copiers, how to estimate the true value? Further, how to incentivize the strategic workers with high accuracy to participate in the crowdsourcing?

We consider a crowdsourcing system consisting of a platform that launches the crowdsourcing campaign, and a set of workers who are connected to the platform via the cloud. We model the crowdsourcing process as a sealed reverse auction. First, the platform publicizes a set of tasks, and each is associated with an accuracy requirement, which is the least confidence to discover the truth. The workers who are interested in performing the tasks can bid with the data for participating. Then, the platform executes the truth discovery for each task. Meanwhile, the accuracy of each worker is estimated in the truth discovery process. Finally, the platform selects a subset of workers as winners, and determines the payment to winners based on the bid price and accuracy of workers. The whole process is illustrated by Fig.1.

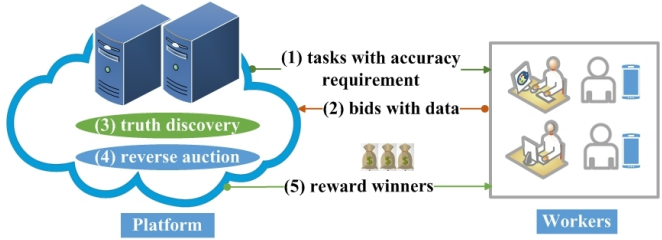

Fig. 1 Reverse auction based crowdsourcing process

We aim to present an *Incentive Mechanism* for *Crowdsourcing with Copiers* ($IMC^2$), which is a two stage incentive mechanism, consisting of truth discovery stage and reverse auction stage. In truth discovery stage, $IMC^2$ performs the *Dependence and Accuracy based Truth Estimation* (*DATE*), and returns the accuracy of workers at the same time. In the stage of reverse auction, $IMC^2$ selects the winners and determinate the payment to the workers.

The problem of designing truthful incentive mechanism for the truth discovery in crowdsourcing with copiers is very challenging. First, we do not know how workers obtain their data, so we need to detect copiers from a snapshot of data. It is challenging to detect the copiers because submitting the same data with others does not imply the copying behavior directly. Second, if any two workers submit the same data, it is not obvious which one is the copier only based on a snapshot of data. This means that we should calculate the dependence with directions. Furthermore, the effective method of accuracy calculation is needed for the copiers since the copiers may contribute to the truth discovery by submitting the combination of the manual data and copied data, or submitting the copied data after verification. Finally, the workers may take a strategic behavior by submitting dishonest bid price to maximize their utility.

The main contributions of this paper are as follows:
- To the best of our knowledge, we are the first to design the incentive mechanism, which stimulate the strategic workers to reach the least confidence for truth discovery in the crowdsourcing.
- We propose a truth discovery algorithm as a component of the incentive mechanism for the crowdsourcing with copiers. Our truth discovery algorithm considers both the dependence and accuracy of workers. Further, we extend our truth discovery algorithm to the general cases, where the truth can have multiple presentations and the distribution of false values is nonuniform.
- We model the *Social Optimization Accuracy Coverage* (*SOAC*) problem, and design a reverse auction mechanism to solve the *SOAC* problem. We show that the designed mechanism satisfies the desirable properties of computational efficiency, individual rationality, truthfulness, and guaranteed approximation.

The rest of the paper is organized as follows. Section II models the *SOAC* problem and the dependence of workers, and lists some desirable properties. Section III presents the detailed design of truth discovery. Section IV extends the truth discovery to the general cases. Section V presents the detailed design of our reverse auction. Section VI presents the detailed analysis of designed incentive mechanism. Performance evaluation is presented in Section VII. We review the state-of-art research in Section VIII, and conclude this paper in Section IX.

II. SYSTEM MODEL

In this section, we model the truth discovery in the crowdsourcing system as a reverse auction. Then we define the dependence model of the workers. At the end of this section, we present some desirable properties.

*A. Reverse Auction Model*

We consider a crowdsourcing system consisting of a platform and a set $W = \{1, 2, ..., n\}$ of $n$ workers, who are interested in performing the crowdsourcing tasks. The platform resides in the cloud. The platform publicizes a set $T = \{t_1, t_2, ..., t_m\}$ of $m$ tasks, and wants to discover the truth for each task from the data submitted by the workers. Each task $t_j \in T$ is associated with the accuracy requirement $\Theta^j$, which is the least confidence to discover the truth for $t_j$. Let $\Theta = (\Theta^1, \Theta^2, ..., \Theta^m)$ be the accuracy requirement profile for all tasks. Without loss of generality, we consider that each task $t_j$ has ($num^j + 1$) different answers. In other words, there are one true value and $num^j$ false values.

Each worker $i \in W$ submits a triple $B_i = (T_i, b_i, D_i)$, where $T_i$ is the task set he/she is willing to perform, and $b_i$ is his/her bid price that worker $i$ wants to charge for performing $T_i$. Each $T_i$ is associated with the cost $c_i$, which is the private information and known only to worker $i$. Different from most existing crowdsourcing systems [33, 34, 38, 39], each worker sends his/her data $D_i$ of task set $T_i$ to the platform at the same time. Let $\mathbf{D} = (D_1, D_2, ..., D_n)$ be the data submitted by all workers.

Given the task set $T$, the bid profile $\mathbf{B} = (B_1, B_2, ..., B_n)$, and the accuracy requirement profile $\mathbf{\Theta} = (\Theta^1, \Theta^2, ..., \Theta^m)$, the platform calculates the estimated truth $\mathbf{et} = (et^1, et^2, ..., et^m)$ for each task, the winner set $S \subseteq W$, and the payment $p_i$ for each winner $i \in S$. We define the utility of any worker $i$ as the difference between the payment and its real cost:

$$u_i = p_i - c_i . \tag{1}$$

Since we consider the workers are selfish and rational individuals, each worker can behave strategically by submitting a dishonest bid price to maximize its utility.

The utility of the platform is:

$$u_0 = V(S) - \sum_{i \in S} p_i \tag{2}$$

where $V(S)$ is the value of the platform obtained if all of the tasks can be completed by the workers in $S$ with accuracy no less than the accuracy requirement.

We define the social welfare as the total utility of the platform and all workers:

$$u_{social} = u_0 + \sum_{i \in W} u_i = V(S) - \sum_{i \in S} c_i \tag{3}$$

We consider an incentive mechanism $\mathcal{M}(e, f, p)$ consisting of an truth estimation function $e(\cdot)$, an winner selection function $f(\cdot)$, and a payment function $p(\cdot)$. The function $e(\cdot)$ estimates the truth $\mathbf{et}$ for all tasks, and returns an accuracy matrix $\mathbf{A} = \{A_i^j\}_{n \times m}$, where $A_i^j$ is the accuracy of worker $i$ for task $t_j$ for any $i \in W, t_j \in T$. The function $f(\cdot)$ outputs the subset of workers $S \subseteq W$. The function $p(\cdot)$ returns a vector $\mathbf{p} = (p_1, p_2, ..., p_n)$ of payments to all winners.

The objective of our incentive mechanism is maximizing the social welfare such that each of tasks in $T$ can be completed with accuracy no less than the accuracy requirement.

Note that the problem of maximizing the social welfare is equivalent to the problem of minimizing the social cost (total cost of winners) since the value of $V(S)$ is constant under the accuracy constraint. We refer this problem as the *Social Optimization Accuracy Coverage* (*SOAC*) problem, which can be formulated as follows:

**Objective:** Minimize $\sum_{i \in S} c_i \cdot x_i$ (4)

**Subject to:** $\sum_{i \in W} A_i^j \cdot x_i \geq \Theta^j, \forall t_j \in T$ (5)

$x_i \in \{0,1\}, \forall i \in W$ (6)

where $x_i$ is the binary variable for each worker $i \in W$. Let $x_i = 1$ if $i$ is a winner; otherwise, $x_i = 0$.

The constraint (5) represents the accuracy coverage for each task $t_j \in T$, which ensures that the total accuracy of all the winners for this task is no less than the accuracy requirement $\Theta_j$.

### B. Dependence Model of Workers

Different from most existing truth discovery methods, we take the dependence of workers into consideration in order to reduce the impact of copiers on truth estimation. We define the dependence of workers in Definition 1.

**Definition 1.** (Dependence of workers) *We say that there exists a dependence between any two workers $i$ and $i'$ if they derive the same part of their data directly from the other worker (can be one of $i$ and $i'$).*

An independent worker provides all values independently. It may provide some erroneous values because of incorrect knowledge of the real world, mis-spelling, etc. We use $W_i \perp W_{i'}$ to represent that workers $i$ and $i'$ are independent.

A copier copies a part (or all) of data from other workers (independent workers or copiers). Let $r$ be the probability that a value provided by a copier is copied. The copier can copy from multiple workers by union, intersection, etc. In addition, a copier may revise some of the copied values or add additional values. Such revised and added values are considered as independent contributions of the copier. We consider the case when any two workers $i$ and $i'$ are dependent, denoting $i$ depending on $i'$ by $i \to i'$, and $i'$ depending on $i$ by $i' \to i$, respectively.

To make the computation tractable, we assume that the dependence of workers satisfies the following properties:

- *Independent copying*: The dependence of any pair of workers is independent of the dependence of any other pair of workers.
- *No loop dependence*: The dependence relationship between workers is non-transitive.
- *Uniform false-value distribution*: For each task, there are multiple false values in the underlying domain, and an independent worker has the same probability of providing each of them (we will remove this assumption in the general cases considered in Section IV).

### C. Desirable Properties

Our objective is to design an incentive mechanism satisfying the following desirable properties:

- **Computational efficiency:** An incentive mechanism is computationally efficient if the truth estimation $\mathbf{et}$, the winner set $S$, and the payment vector $\mathbf{p}$ can be computed in polynomial time.
- **Individual Rationality:** Each winner will have a non-negative utility while bidding its true cost, i.e. $u_i \geq 0, \forall i \in S$.
- **Truthfulness:** An incentive mechanism is truthful if reporting the true cost is a weakly dominant strategy for all workers. In other words, no worker can improve its utility by submitting a false cost, no matter what others submit.
- **Social Optimization:** The objective is minimizing the social cost. We attempt to find optimal solution or approximation algorithm with low approximation ratio when there is no optimal solution terminated in polynomial time. For the latter, the approximation ratio is the ratio between approximation solution and the optimal solution.

The importance of the first two properties is obvious, because they together assure the feasibility of the incentive mechanism. The third property is indispensable for guaranteeing the compatibility. Being truthful, the incentive mechanism can eliminate the fear of market manipulation and the overhead of strategizing over others for the workers. The last property guarantees that the incentive mechanism can have a guaranteed approximation ratio to close to the optimal social cost.

## III. Truth Discovery

In this section, we present our truth discovery algorithm *DATE*, which discovers the true values from conflicting information provided by multiple workers. *DATE* performs the following three steps (the details will be shown in subsection *A*, *B*, and *C*, respectively) illustrated by Fig.2 iteratively until the estimated truth does not change or the number of iteration exceed the maximum number of iterations $\varphi$.

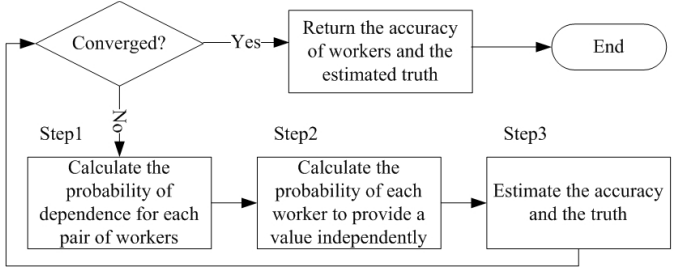

Fig. 2 Workflow of *DATE*

### A. Dependence Between the Workers

We consider that there are two types of workers: independent workers and copiers. For any pair of workers $i \in W, i' \in W, i \neq i'$, we apply Bayesian analysis to compute the probability that $i$ and $i'$ are dependent given the observation of data set **D**. For this purpose, we need to compute the probability of the observed data, conditioned on the dependence or independence of these two workers.

We are interested in three sets of tasks: $T^s$, the set of tasks on which $i$ and $i'$ provide the same true value; $T^f$, the set of tasks on which they provide the same false values; $T^d$, the set of tasks on which they provide different values.

Note that the true value can be obtained through the voting mechanism on data set **D** for each task initially. In the following iterations, the true value will be determined based on the estimated truth **et**.

Intuitively, two independent workers providing the same false value is a rare event; thus, if we fix $T^s \cup T^f$ and $T^d$, the more common false values that $i$ and $i'$ provide, the more likely that they are dependent. On the other hand, if we fix $T^s$ and $T^f$, the fewer tasks on which $i$ and $i'$ provide different values, the more likely that they are dependent. We next describe how we compute conditional probability of **D** based on this idea.

We first consider the situation where the two workers $i$ and $i'$ are independent. Since there is only one true value, the probability that $i$ and $i'$ provide the same true value for task $t_j$, denoted by $P_s^j$ for convenience, is

$$P_s^j = P(t_j \in T^s \mid i \perp i') = A_i^j \cdot A_{i'}^j \tag{7}$$

where $A_i^j$ and $A_{i'}^j$ are the accuracy of $i$ and $i'$ for task $t_j$ respectively. We set $A_i^j = \varepsilon$ for $\forall i \in W, \forall t_j \in T$, $\varepsilon \in (0,1)$ be the default values initially, and iteratively refine them by computing the estimated values in later rounds of *DATE*.

Based on the assumption of *uniform false-value distribution* made in subsection II-B, any independent worker has the same probability of providing each false value of task $t_j$. Thus the probability that any worker $i$ provides a false value for task $t_j$ is $\frac{1-A_i^j}{num^j}$. Thus, the probability that $i$ and $i'$ provides the same false value for task $t_j$, denoted by $P_f^j$, is

$$\begin{aligned}P_f^j &= P(t_j \in T^f \mid i \perp i') \\ &= num^j \cdot \frac{1-A_i^j}{num^j} \cdot \frac{1-A_{i'}^j}{num^j} = \frac{(1-A_i^j)\cdot(1-A_{i'}^j)}{num^j}\end{aligned} \tag{8}$$

Then, the probability that $i$ and $i'$ provide different values on task $t_j$, denoted by $P_d^j$, is

$$P_d^j = P(t_j \in T^d \mid i \perp i') = 1 - P_s^j - P_f^j \tag{9}$$

Thus, the conditional probability of observing **D** is

$$P(\mathbf{D} \mid i \perp i') = \prod_{t_j \in T^s} P_s^j \cdot \prod_{t_j \in T^f} P_f^j \cdot \prod_{t_j \in T^d} P_d^j \tag{10}$$

We next consider the situation where $i$ and $i'$ have the dependence relationship. There are two cases where $i$ and $i'$ provide the same value for the task $t_j$. First, with probability $r$, one copies the value from the other (there assumes $i$ copies from $i'$) and so the value is true with probability $A_{i'}^j$ and false with probability $1-A_{i'}^j$. Second, with probability $1-r$, the two workers provide the value independently, and so the probability of being true or false is the same as that in the situation where $i$ and $i'$ are independent. Thus, when $i$ copies from $i'$ (similar for $i'$ copying from $i$), we have

$$P(t_j \in T^s \mid i \to i') = A_{i'}^j \cdot r + P_s^j \cdot (1-r) \tag{11}$$

$$P(t_j \in T^f \mid i \to i') = (1-A_{i'}^j) \cdot r + P_f^j \cdot (1-r), \tag{12}$$

Finally, the probability that $i$ and $i'$ provide the different values on task $t_j$ is the probability that $i$ provides a value independently, and the value differs from that provided by $i'$:

$$P(t_j \in T^d \mid i \to i') = P_d^j \cdot (1-r) \tag{13}$$

Thus, the conditional probability of observing **D** is

$$\begin{aligned}&P(\mathbf{D} \mid i \to i') \\ &= \prod_{t_j \in T^s}[A_{i'}^j \cdot r + P_s^j \cdot (1-r)] \\ &\cdot \prod_{t_j \in T^f}[(1-A_{i'}^j)\cdot r + P_f^j \cdot (1-r)] \cdot \prod_{t_j \in T^d}[P_d^j \cdot (1-r)]\end{aligned} \tag{14}$$

We compute $P(i \to i' \mid \mathbf{D})$ accordingly:

$$P(i \to i' | \mathbf{D})$$
$$= \frac{P(\mathbf{D} | i \to i')P(i \to i')}{P(\mathbf{D} | i \to i')P(i \to i') + P(\mathbf{D} | i \perp i')P(i \perp i')}$$
$$= [1 + (\frac{1-\alpha}{\alpha}) \cdot \prod_{t_j \in T^s} \frac{P_s^j}{A_{i'}^j \cdot r + P_s^j \cdot (1-r)} \quad (15)$$
$$\cdot \prod_{t_j \in T^f} \frac{P_f^j}{(1 - A_{i'}^j) \cdot r + P_f^j \cdot (1-r)} \cdot (\frac{1}{1-r})^{|T^d|}]^{-1}$$

where $P(i \to i')$ is the a priori probability that worker $i$ and $i'$ are dependent. Let $P(i \to i') = \alpha, P(i \perp i') = (1-\alpha), 0 < \alpha < 1$ be the default values for every pair of workers initially, and iteratively refine them by computing the estimated values in later rounds of *DATE*.

Note that the probability of $i$ and $i'$ providing the same true or false value is different with different directions of dependence. By applying the Bayesian rule, we can compute the probabilities of $i \perp i'$, $i \to i'$, and $i' \to i$ for any pair of workers $i$ and $i'$.

*B. Probability of Providing the Value Independently*

We have described how to detect any pair of workers are dependent. However, if a worker copies from another, it is possible that it provides some of the values independently, and it would be inappropriate to ignore the contribution of these values. Thus, we describe how to obtain the probability that any worker provides the value independently in this subsection.

Note that the probability of dependence calculated by formula (15) is based on the whole data collected. To estimate the truth for each task, we should calculate the probability of providing each possible value independently. Obviously, it would take exponential time to enumerate all possible dependence for each value between all pairs of workers.

To make the calculation scalable, we shall find a method with polynomial time. The basic idea is calculating the probability of providing each possible value $v$ by considering the worker one by one for every task. For convenience, let $D^j$ be the set of values of any task $t_j \in T$. Let $W_v^j$ be the set of workers who provide value $v$ for any task $t_j \in T$. The goal is to calculate the probability of any worker $i$ to provide each possible value $v$ of any task $t_j$ independently, denoted as $I_v^j(i)$.

For each task $t_j \in T$ and $v \in D^j$, we denote an ordered set $\overline{W_v^j}$, and put the workers in $W_v^j$ into $\overline{W_v^j}$ one by one. For each worker $i \in W_v^j, i \notin \overline{W_v^j}$, we compute the probability for $i$ based on the dependence on the workers in $\overline{W_v^j}$. This method is not precise because if any worker $i$ depends only on workers in $W_v^j \setminus \overline{W_v^j}$ but some of those workers in $W_v^j \setminus \overline{W_v^j}$ depend on the workers in $\overline{W_v^j}$, our estimation still consider that the worker $i$ provides the value independently.

To minimize such error, we wish that both the probability that worker $i$ depends on the workers in $W_v^j \setminus \overline{W_v^j}$ and the probability that the workers in $W_v^j \setminus \overline{W_v^j}$ depend on the workers in $\overline{W_v^j}$ be the lowest. Thus, we take a greedy algorithm and consider workers in such an order: In the first round, we select a worker $i_0 \in W_v^j$ that is associated with the highest dependence probability, and make the worker as the first one in ordered set $\overline{W_v^j}$; In the later rounds, we select a worker that has the maximal dependence probability on one of the previously selected workers. And this process ends when all workers are considered.

Thus the probability that the worker $i_0$ provides value $v$ of task $t_j$ independently is

$$I_v^j(i_0) \leftarrow \prod_{i' \in \overline{W_v^j}} (1 - r \cdot P(i_0 \to i' | \mathbf{D})) \quad (16)$$

*C. Accuracy and Truth Estimation*

We next consider how to compute the accuracy of a worker. A straightforward way is to compute the fraction of true values provided by the worker. However, we do not know which the true values are exactly. We instead compute the accuracy of a worker as the average probability of its values for any task $t_j \in T$ being true.

Formally, $D_i^j$ be the set of values of any task $t_j \in T$ provided by worker $i$. For each $v \in D_i^j$, we denote $P^j(v)$ as the probability that $v$ is true for any task $t_j \in T$. We compute $A_i^j$ as follows.

$$A_i^j = \frac{\sum_{v \in D_i^j} P^j(v)}{|D_i^j|}. \quad (17)$$

Now we need a way to compute $P^j(v)$. Intuitively, the computation should consider both how many workers provide the value and the accuracies of those workers. We apply a Bayesian analysis again.

We start with the case where all workers are independent. Consider a task $t_j \in T$, for the observation $D^j$ provided by each worker $i \in W^j$, where $W^j$ is the set of workers who perform task $t_j$, we first compute the probability of $D^j$ conditioned on $z$ being true. This probability represents that the workers in $W_v^j$ provide the true value and the other workers in $W^j$ provide one of the false values.

$$P(D^j | v \text{ is true}) = \prod_{i \in W_v^j} A_i^j \cdot \prod_{i \in W^j \setminus W_v^j} \frac{1 - A_i^j}{num^j} \quad (18)$$

Among the values in $D^j$, there is one and only one true value. Applying the a-priori belief of each value being true is the same, denoted by $\beta$. We then have

$$P(D^j) = \sum_{v \in D^j} (\beta \cdot \prod_{i \in W_v^j} A_i^j \cdot \prod_{i \in W^j \setminus W_v^j} \frac{1 - A_i^j}{num^j}) \quad (19)$$

Applying the Bayesian rule, we have

**Algorithm 1: DATE**

**Input:** worker set $W$, task set $T$, data set $\mathbf{D}$, copy probability $r$, initial accuracy $\varepsilon$, priori probability of dependence $\alpha$, maximum number of iterations $\varphi$

**Output:** estimated truth **et**, accuracy matrix $\mathbf{A}$

1: **for each** $i \in W$ **do**
2:     **for each** $t_j \in T$ **do** $A_i^j \leftarrow \varepsilon$;
3:     **for each** $i' \in W$, s.t. $i' \neq i$ **do**
4:        $P(i \to i') \leftarrow \alpha, P(i \perp i') \leftarrow (1-\alpha)$;
5:     **end for**
6: **end**
7: $\mathcal{K} \leftarrow 0$;
8: **while et** $\neq$ **et'** and $\mathcal{K} \leq \varphi$ **do**
9:     **for each** $t_j \in T$ **do**
10:       **for each** $v \in D^j$ **do** $\overline{W_z^j} \leftarrow \varnothing$;
11:     **end for**
12:     **et** $\leftarrow$ **et'**;
       //Step1: Calculate the probability of dependence
13:     calculate $P(i \to i'|\mathbf{D})$ for every pair of workers $i, i' \in W, i \neq i'$ through formula (15) with **et** and $\mathbf{A}$;
       // Step2: Calculate the probability to provide a value independently
14:     **for each** $t_j \in T$ **do**
15:       **for each** $v \in D^j$ **do**
16:         $i_0 \leftarrow \arg\min_{i:i,i' \in W_v^j, i \neq i'} P(i \to i'|\mathbf{D}) + P(i' \to i|\mathbf{D})$;
17:         $\overline{W_v^j} \leftarrow \{i_0\}$;
18:         **while** $|\overline{W_v^j}| \neq |W_v^j|$ **do**
19:            $i_0 \leftarrow \arg\max_{i:i \in W_v^j \setminus \overline{W_v^j}, i' \in \overline{W_v^j}, i \neq i'} P(i \to i'|\mathbf{D})$;
20:            $I_v^j(i_0) \leftarrow \prod_{i' \in \overline{W_v^j}} (1 - r \cdot P(i_0 \to i'|\mathbf{D}))$;
21:            $\overline{W_v^j} \leftarrow \overline{W_v^j} \cup \{i_0\}$;
22:         **end while**
         // Step3: Estimate the accuracy and the truth
23:         $P^j(v) \leftarrow \dfrac{\prod_{i \in W_v^j} \dfrac{num^j \cdot A_i^j}{1-A_i^j}}{\sum_{v' \in D^j} \prod_{i \in W_{v'}^j} \dfrac{num^j \cdot A_i^j}{1-A_i^j}}$
24:       **end for**
25:       **for each** $i \in W$, s.t. $t_j \in T_i$ **do**
26:         $A_i^j \leftarrow \dfrac{\sum_{v \in D_i^j} P^j(v)}{|D_i^j|}$;
27:       **end for**
28:       $et^j \leftarrow \arg\max_{v \in D^j} \sum_{i \in W_v^j} A_i^j \cdot I_v^j(i)$;
29:     **end for**
30:     $\mathcal{K} \leftarrow \mathcal{K} + 1$;
31: **end while**

$$P^j(v) = P(v \text{ is true} | D^j) = \frac{P(D^j | v \text{ is true})P(v \text{ is true})}{P(D^j)}$$

$$= \frac{\prod_{i \in W_v^j} \dfrac{num^j \cdot A_i^j}{1-A_i^j}}{\sum_{v' \in D^j} \prod_{i \in W_{v'}^j} \dfrac{num^j \cdot A_i^j}{1-A_i^j}} \quad (20)$$

For the truth discovery, if a worker $i$ copies a value $v$ from other workers, we should ignore $i$ when considering $v$ as the truth. Thus, we adopt $\sum_{i \in W_v^j} A_i^j \cdot I_v^j(i)$ as the support counts of value $v$ for any task $t_j \in T$, and find the value with the maximal support counts in $D^j$. In the last round of *DATE*, the value with highest support counts is the final estimated truth.

The whole process of *DATE* is illustrated in Algorithm 1.

## IV. TRUTH DISCOVERY FOR GENERAL CASES

### A. Discover the Truth with Multiple Presentations

In some scenarios, part of workers may submit certain values in abbreviations, missing or incorrect spelling. These values mean the same thing, but without distinction, they will be treated as different values. For example, we should treat *IT* and *Information Technology* as the same value. Therefore, we need a method to calculate the similarities between different values.

For any task $t_j \in T$, if any value $v' \in D^j$ is similar to another value $v \in D^j$, Intuitively, the workers that support for $v'$ also implicitly support for $v$.

Formally, we denote the similarity between $v$ and $v'$ as $sim(v,v') \in [0,1]$, which can be converted to the similarity of word vectors[25], then computed by *Euclidean Distance* [21], *Pearson Correlation* [22], *Asymmetric Similarity* [23], *Cosine Similarity* [24], etc.

After computing the support counts of each value for any task $t_j \in T$, we adjust it by considering the similarities between them as follows:

$$\sum_{i \in W_v^j} A_i^j \cdot I_v^j(i) + \rho \cdot \sum_{v' \in D^j, v' \neq v} \sum_{i' \in W^j \setminus W_v^j} A_{i'}^j \cdot I_v^j(i') \cdot sim(v,v') \quad (21)$$

where $\rho \in [0,1]$ is a parameter controlling the influence of similar values.

We then use the adjusted support counts for truth estimation (line 28 of Algorithm 1).

### B. Nonuniform false-value distribution

In Section III, we estimate the truth with the assumption of *uniform false-value distribution* made in subsection II-B. However, the false values of a task may not be uniformly distributed. For example, in the minds of most people, Australia's capital is Sydney, but in fact, Canberra is its capital. The probability of false value of Sydney will larger than other false values.

We define $f(h), h \in [0,1]$, as the percentage of false values whose distribution probability is $h$; thus, $\int_0^1 f(h)dh = 1$. Then,

the probability that two false-value providers provide the same value is $\int_0^1 h^2 f(h)dh$ instead of $(\frac{1}{num^j})^2 \cdot num^j = \frac{1}{num^j}$. Accordingly, we revise formula (8) as

$$P_f^j = P(t_j \in T^f \mid i \perp i') = (1-A_i^j) \cdot (1-A_i^j) \cdot \int_0^1 h^2 f(h)dh \quad (22)$$

Similarly, we need to revise formula (18) as follows:

$$P(D^j \mid v \text{ is true})$$
$$= \Pi_{i \in W_v^j} A_i^j \cdot \Pi_{i \in W^j \setminus W_v^j} (1-A_i^j \cdot e^{\int_0^1 \ln df(h) \cdot |W^j \setminus W_v^j|}) \quad (23)$$

## V. REVERSE AUCTION DESIGN

First of all, we attempt to find an optimal algorithm for the *SOAC* problem presented in equation (2)∼(4). Unfortunately, as the following theorem shows, the *SOAC* problem is NP-hard.

**Theorem 1.** *The SOAC problem is NP-hard.*

*Proof*: We consider a special case of *SOAC* problem, where the accuracy requirements for all tasks in *T* are the same. Let $\Theta^j$ be sufficiently close to zero $\forall t_j \in T$. This means that, in this special case, any task $t_j \in T$ can be completed upon there is any worker $i \in W$ with $A_i^j > 0$. In this way, the problem can be simplified as selecting a subset $S \subseteq W$ with minimum total cost such that the workers in *S* can perform every task in *T*. Since each worker can bid for a subset of *T* with a cost, this special problem is actually an instance of the <u>Weighted Set Cover</u> (*WSC*) problem, which can be formulated as follows:

**Objective:** Minimize $\sum_{i \in S} b_i$ (24)

**Subject to:** $\sum_{i \in S} A_i^j > 0, \ \forall t_j \in T$ (25)

Since the *WSC* problem is a well-known NP-hard problem, the *SOTD* problem is NP-hard. ∎

Since the *SOAC* problem is NP-hard, it is impossible to compute the winner set with minimum social cost in polynomial time unless P=NP. In fact, there is no $(1-\varepsilon)\ln n$ approximate polynomial time algorithm for *WSC* problem [26]. In addition, we cannot use the off-the-shelf *VCG* mechanism [27] since the truthfulness of *VCG* mechanism requires that the social cost is exactly minimized. We design our reverse auction, which follows a greedy approach. Illustrated in Algorithm 2, our reverse auction consists of winner selection phase and payment determination phase.

In the winner selection phase, the workers are sorted according to the effective accuracy unit cost, which is defined as $\frac{b_i}{\sum_{t_j \in T_i} \min\{\Theta^j{}', A_i^j\}}$ for any worker $i \in W$. In each iteration of the winner selection phase, we select the worker with minimum effective accuracy unit cost over the unselected worker set $W \setminus S$ as the winner until the winners' accuracy can cover the accuracy requirement for each task in *T*.

In payment determination phase, for each winner $i \in S$, we execute the winner selection phase over $W \setminus \{i\}$, and the winner set is denoted as $S'$. We compute the maximum price that worker *i* can be selected instead of each worker in $S'$. We will prove that this price is a critical payment for worker *i* later.

---

**Algorithm 2: Reverse Auction**

**Input:** task set *T*, bid profile *B*, worker set *W*, accuracy requirement profile $\Theta$, accuracy matrix **A**

**Output:** winner set *S*, payment **p**

//Winner Selection Phase
1: $S \leftarrow \varnothing, \Theta' \leftarrow \Theta$;
2: **while** $\sum_{t_j \in T} \Theta^j{}' \neq 0$ **do**
3: $\quad i \leftarrow \arg\min_{k \in W \setminus S} \frac{b_k}{\sum_{t_j \in T_k} \min\{\Theta^j{}', A_k^j\}}$;
4: $\quad S \leftarrow S \cup \{i\}$;
5: $\quad$ **for each** $t_j \in T_i$ **do**
6: $\quad\quad \Theta^j{}' \leftarrow \Theta^j{}' - \min\{\Theta^j{}', A_i^j\}$;
7: $\quad$ **end for**
8: **end while**
//Payment Determination Phase
9: **for each** $i \in W$ **do** $p_i \leftarrow 0$;
10: **for each** $i \in S$ **do**
11: $\quad W' \leftarrow W \setminus \{i\}, S' \leftarrow \varnothing, \Theta'' \leftarrow \Theta$;
12: $\quad$ **while** $\sum_{t_j \in T} \Theta^j{}'' \neq 0$ **do**
13: $\quad\quad i_k \leftarrow \arg\min_{k \in W' \setminus S'} \frac{b_k}{\sum_{t_j \in T_k} \min\{\Theta^j{}'', A_k^j\}}$;
14: $\quad\quad S' \leftarrow S' \cup \{i_k\}$;
15: $\quad\quad p_i \leftarrow \max\{p_i, \frac{\sum_{t_j \in T_i} \min\{\Theta^j{}'', A_i^j\}}{\sum_{t_j \in T_{i_k}} \min\{\Theta^j{}'', A_{i_k}^j\}} b_{i_k}\}$;
16: $\quad\quad$ **for each** $t_j \in T_{i_k}$ **do**
17: $\quad\quad\quad \Theta^j{}'' \leftarrow \Theta^j{}'' - \min\{\Theta^j{}'', A_{i_k}^j\}$;
18: $\quad\quad$ **end for**
19: $\quad$ **end while**
20: **end for**

---

## VI. MECHANISM ANALYSIS

In the following, we present the theoretical analysis, demonstrating that $IMC^2$ can achieve the desired properties of computational efficiency, individual rationality, truthfulness, and low approximation ratio.

**Lemma 1.** $IMC^2$ *is computationally efficient.*

*Proof:* The running time of Algorithm 1 is dominated by the while loop for sorting the workers in $\overline{W_z^j}$ (line 18-22), which takes $O(n^2)$ since there are at most *n* workers in $W_z^j$. Since *DATE* executes the sorting for each value of each task, and the maximal number iteration is $\varphi$, *DATE* is bounded by $O(\varphi n^2 m \max_{j=1,2,...,m} \{num^j\})$.

For Algorithm 2, finding the worker with minimum effective accuracy unit cost takes $O(nm)$, where computing the value of $\sum_{t_j \in T_k} \min\{\Theta^{j'}, A_k^j\}$ takes $O(m)$. Hence, the while-loop (line 2-8) takes $O(n^2m)$. In each iteration of the for-loop (line 10-20), a process similar to line 2-8 is executed. Hence the time complexity of the whole reverse auction is dominated by this for-loop, which is bounded by $O(n^3m)$. ■

**Lemma 2.** $IMC^2$ is individually rational.

*Proof:* Let $i_k$ be worker $i$'s replacement which appears in the $i$th place in the sorting over $W \setminus \{i\}$. Since worker $i_k$ would not be at $i$th place if $i$ is considered, we have $\frac{b_i}{\sum_{t_j \in T_i} \min\{\Theta^{j'}, A_i^j\}} \leq \frac{b_{i_k}}{\sum_{t_j \in T_{i_k}} \min\{\Theta^{j'}, A_{i_k}^j\}}$. Hence

$$b_i \leq \frac{\sum_{t_j \in T_i} \min\{\Theta^{j'}, A_i^j\}}{\sum_{t_j \in T_{i_k}} \min\{\Theta^{j'}, A_{i_k}^j\}} b_{i_k} = \frac{\sum_{t_j \in T_i} \min\{\Theta^{j''}, A_i^j\}}{\sum_{t_j \in T_{i_k}} \min\{\Theta^{j''}, A_{i_k}^j\}} b_{i_k},$$

where the equality relies on the observation that $\Theta^{j'} = \Theta^{j''}$ for every $k \leq i$, which is due to the fact that $S = S'$ for every $k \leq i$. This is sufficient to guarantee $b_i \leq \max_{k \in W \setminus S'} \frac{\sum_{t_j \in T_i} \min\{\Theta^{j''}, A_i^j\}}{\sum_{t_j \in T_{i_k}} \min\{\Theta^{j''}, A_{i_k}^j\}} b_{i_k} = p_i$ ■

Before analyzing the truthfulness of $IMC^2$, we first introduce the Myerson's Theorem [28].

**Theorem 2.** *([29, Theorem 2.1]) An auction mechanism is truthful if and only if:*

- *The selection rule is monotone: If worker $i$ wins the auction by bidding $b_i$, it also wins by bidding $b_i' < b_i$;*
- *Each winner is paid the critical value: Worker $i$ would not win the auction if it bids higher than this value.*

**Lemma 3.** $IMC^2$ is truthful.

*Proof:* Based on Theorem 2, it suffices to prove that the selection rule of $IMC^2$ is monotone and the payment $p_i$ for each $i$ is the critical value. The monotonicity of the selection rule is obvious as bidding a lower price cannot push worker $i$ backwards in the sorting.

We next show that $p_i$ is the critical value for worker $i$ in the sense that bidding higher $p_i$ could prevent worker $i$ from winning the auction. Note that $p_i = \max_{k \in \{1,...,e\}} \frac{\sum_{t_j \in T_i} \min\{\Theta^{j''}, A_i^j\}}{\sum_{t_j \in T_{i_k}} \min\{\Theta^{j''}, A_{i_k}^j\}} b_{i_k}$. If worker $i$ bids $b_i \geq p_i$, it will be placed after $e$ since $b_i \geq \frac{\sum_{t_j \in T_i} \min\{\Theta^{j''}, A_i^j\}}{\sum_{t_j \in T_{i_e}} \min\{\Theta^{j''}, A_{i_e}^j\}} b_{i_e}$ implies $\frac{b_i}{\sum_{t_j \in T_i} \min\{\Theta^{j''}, A_i^j\}} \geq \frac{b_{i_e}}{\sum_{t_j \in T_{i_e}} \min\{\Theta^{j''}, A_{i_e}^j\}}$. Hence, worker $i$ would not win the auction because the first $e$ workers have met the accuracy requirement for each task in $T$. ■

Then, we provide our analysis about the approximation ratio of $IMC^2$ using the dual fitting method [30]. The normalized primal linear program **P** has been formulated in equation (4)～(6). The dual program **D** is formulated in equation (26)～(29).

$$\mathbf{D}: \max \sum_{t_j \in T} \Theta^j y_j - \sum_{i \in W} z_i \quad (26)$$

$$\text{s.t. } \sum_{t_j \in T_i} (A_i^j y_j) - z_i \leq b_i, \quad \forall i \in W \quad (27)$$

$$y_j \geq 0, \quad \forall t_j \in T \quad (28)$$

$$z_i \geq 0, \quad \forall i \in W \quad (29)$$

We define any task $t_j \in T$ as alive at any iteration in winner selection phase if its accuracy requirement is not fully satisfied. We define that task $t_j$ is covered by $T_i$ if $t_j \in T_i$ and $t_j$ is alive when worker $i$ is selected. The coverage relationship is represented as $t_j \prec T_i$. Moreover, we define the minimum accuracy as $\Delta v$. Suppose when worker $i$ is selected, the residual accuracy requirement profile is $\{\Theta^{1*}, \Theta^{2*}, ..., \Theta^{m*}\}$ and $T_i$ is the $i_j$th set that covers $t_j$, the corresponding normalized effective accuracy unit cost in terms of unit accuracy can be represented in equation (30):

$$w(t_j, i_j) = \frac{b_i \Delta v}{\sum_{t_j \in T_i} \min\{\Theta^{j*}, A_i^j\}} \quad (30)$$

We assume that $t_j$ is covered by $h_j$ sets. Then we have $w(t_j, 1) \leq ... \leq w(t_j, h_j)$. We then define two constants $\Omega = \frac{1}{\Delta v} \sum_{t_j \in T} \Theta^j$ and $\varepsilon = \max A_i^j \cdot |T_i| \cdot b_i, i \in W, t_j \in T$.

**Lemma 4:** *The following pairs $(y_j, z_i), t_j \in T, i \in W$ are feasible to the dual program **D**.*

$$y_j = \frac{w(t_j, h_j)}{2\varepsilon H_n \Delta v}, \forall t_j \in T,$$

$$z_i = \begin{cases} \frac{\sum_{t_j \prec T_i} \left(\min\{\Theta^{j*}, A_i^j\}(w(t_j, h_j) - w(t_j, i_j))\right)}{2\varepsilon H_\Omega \Delta v}, & i \in S \\ 0, & i \notin S \end{cases}$$

where $H_n = 1 + \frac{1}{2} + ... + \frac{1}{n}$, $H_\Omega = 1 + \frac{1}{2} + ... + \frac{1}{\Omega}$.

*Proof:* Suppose for any worker $i \in W$, there are $s_i$ tasks in $T_i$. We reorder these tasks in the order in which they are fully covered.

If $i \notin S$, then we have $z_i = 0$. Suppose when the last unit accuracy requirement of $t_j$ is covered, the residual accuracy requirement profile is $\{\Theta^{1+}, \Theta^{2+}, ..., \Theta^{m+}\}$, then the total residual accuracy requirement of alive tasks contained by $T_i$ are represented as $\sum_{k=j}^{s_i}\min\{\Theta^{k+},A_k^j\}$. We have

$$w(t_j,h_j)\le\frac{b_i\Delta v}{\sum_{k=j}^{s_i}\min\{\Theta^{k+},A_i^k\}}$$

Therefore, we have

$$\sum_{j=1}^{S_i}(v_i(t_j)y_j)-z_i\le\sum_{j=1}^{S_i}\frac{v_i(t_j)b_i}{2\varepsilon H_\Omega\sum_{k=j}^{s_i}\min\{\Theta^{k+},A_i^k\}}-0$$

$$\le\frac{b_i}{H_\Omega}\left(1+\frac{1}{2}+\ldots+\frac{1}{\Omega}\right)\le b_i$$

If worker $i\in S$, then we assume that when worker $i$ is selected as a winner, $s_i'$ tasks in $T_i$ already been fully covered. We have

$$\sum_{j=1}^{S_i}(A_i^j y_j)-z_i$$

$$=\frac{\sum_{j=1}^{s_i}(w(t_j,h_j)A_i^j)}{2\varepsilon H_\Omega\Delta v}$$

$$-\frac{\sum_{j=s_i'+1}^{s_i}\min\{\Theta^{j*},A_i^j\}\left(w(t_j,h_j)-w(t_j,i_j)\right)}{2\varepsilon H_\Omega\Delta v}$$

$$=\frac{\sum_{j=1}^{s_i'}(w(t_j,h_j)A_i^j)}{2\varepsilon H_\Omega\Delta v}+\frac{\sum_{j=s_i'+1}^{s_i}\min\{\Theta^{j*},A_i^j\}w(t_j,i_j)}{2\varepsilon H_\Omega\Delta v}$$

$$+\frac{\sum_{j=s_i'+1}^{s_i}A_i^j-\min\{\Theta^{j*},A_i^j\}w(t_j,h_j)}{2\varepsilon H_\Omega\Delta v}$$

$$\le\frac{\sum_{j=1}^{s_i'}(w(t_j,h_j)A_i^j)}{2\varepsilon H_\Omega\Delta v}+\frac{\sum_{j=s_i'+1}^{s_i}\min\{\Theta^{j*},A_i^j\}w(t_j,i_j)}{2\varepsilon H_\Omega\Delta v}$$

$$=\sum_{j=1}^{s_i'}\frac{A_i^j b_i}{2\varepsilon H_\Omega\sum_{k=j}^{s_i}\min\{\Theta^{k*},A_i^k\}}+\frac{b_i}{2\varepsilon H_\Omega}$$

$$\le\frac{b_i}{H_\Omega}\left(\frac{1}{s_i}+\ldots+\frac{1}{s_i-s_i'+1}+1\right)\le b_i$$

Hence, the pairs $(y_j,z_i),t_j\in T,i\in W$ are feasible to the dual program **D**. ∎

**Lemma 5:** *$IMC^2$ can approximate the optimal solution within a factor of $2\varepsilon H_\Omega$.*

*Proof:* By substituting the dual solution given in Lemma 4 into equation (26), we have

$$\sum_{t_j\in T}\Theta^j y_j-\sum_{i\in W}z_i$$

$$=\frac{\sum_{i\in S}\sum_{t_j\prec T_i}\left(\min\{\Theta^{j*},A_i^j\}\left(w(t_j,h_j)-w(t_j,i_j)\right)\right)}{2\varepsilon H_\Omega\Delta v}$$

$$+\frac{\sum_{t_j\in T}\Theta^j w(t_j,h_j)}{2\varepsilon H_\Omega\Delta v}$$

$$=\frac{\sum_{i\in S}\sum_{t_j\prec T_i}\min\{\Theta^{j*},A_i^j\}\dfrac{b_i\Delta v}{\sum_{t_j\in T_i}\min\{\Theta^{j*},A_i^j\}}}{2\varepsilon H_\Omega\Delta v}$$

$$=\frac{\sum_{i\in S}b_i}{2\varepsilon H_\Omega}\le OPT$$ ∎

The above lemmas together prove the following theorem.

**Theorem 3.** *$IMC^2$ is computationally efficient, individually rational, truthful and $2\varepsilon H_\Omega$ approximate.*

## VII. PERFORMANCE EVALUATION

We have conducted simulations to investigate the performance of *$IMC^2$* on the real experience data.

### A. Simulation Setup

We first measure the performance of *DATE*, and compare it with following three bench mark algorithms:

- *MV (Majority Voting* [14]*)*: The truth of each task is the corresponding value that supported by the most workers.
- *ED (Enumerate all workers' Dependence)*: Enumerate all possible dependence for each worker with others when calculating the probability of providing each possible value independently (Step 2 of *DATE*).
- *NC (No Copier)*: Consider all workers are independent. All calculations about dependence are not needed in *NC*. This means that *NC* only includes Step 3 of *DATE*.

We define the *precision* of the truth discovery as $\dfrac{\sum_{t_j\in T}g(et^j=et^{*j})}{|T|}$, where $et^{*j}$ is the real truth of task $t_j$, $g(\cdot)$ is an indicator function (*i.e.*, $g(et^j=et^{*j})=1$ if event $et^j=et^{*j}$ is true; otherwise, $g(et^j=et^{*j})=0$).

Then, we conduct the simulations to evaluate the *Reverse Auction*, and compare it with following algorithms:

- *GA (Greedy Accuracy)*: Each time, *GA* selects the worker with the highest accuracy, and pays the critical value [28] to the winners.
- *GB (Greedy Bid)*: Each time, *GB* selects the worker with the lowest bid, and follows the *Vickrey Auction* [20] payment rule.

For our simulations, we use the data from *Qatar Living Forum* [40] to simulate the crowdsourcing network. The data was collected from survey participants using Qatar Living Forum in 2015. It includes 300 questions, 120 workers and 6000 comments, each comment can be annotated as *"Good"*, *"Bad"* or *"Other"*. The cost of each worker is selected randomly from the auction dataset [41], which contains 5017 bid prices for *Palm Pilot M515 PDA* from eBay workers. The task accuracy requirement of tasks is uniformly over [2, 4]. The default number of tasks and workers are 300 and 120, respectively. The value of each task is uniformly distributed over [5, 8]. In the simulations, we randomly selected 30 workers and set them to be copiers. This means that the data of these workers is copied from the other workers. We will

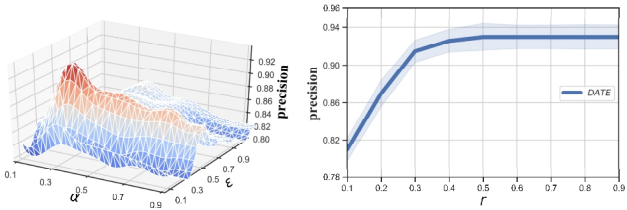

(a) Precision versus $\varepsilon$, $\alpha$   (b) Precision versus $r$

Fig. 3 Impact of parameters on *precision*

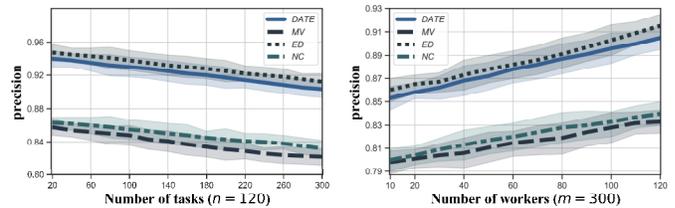

(a) Precision versus tasks   (b) Precision versus workers

Fig. 4 Precision with different number of tasks and workers

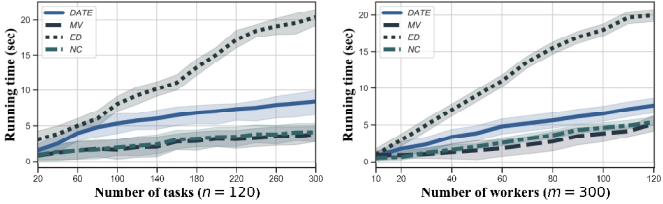

(a) Running time versus tasks   (b) Running time versus workers

Fig. 5 Running time of *DATE*

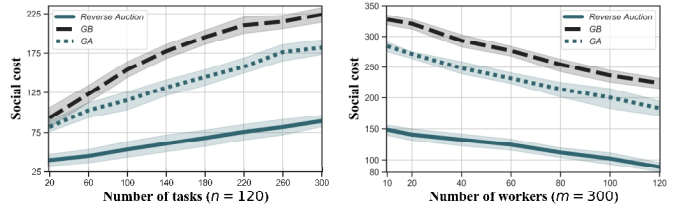

(a) Social cost versus tasks   (b) Social cost versus workers

Fig. 6 Social cost

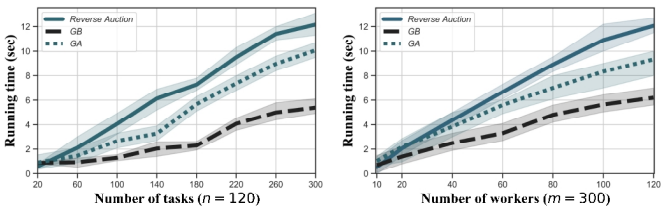

(a) Running time versus tasks   (b) Running time versus workers

Fig. 7 Running time of *Reverse Auction*

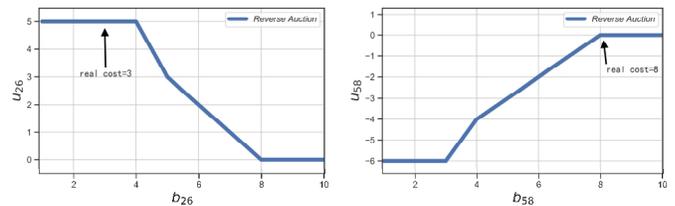

(a) Utility of user with ID=26 (winner)   (b) Utility of user with ID=58 (loser)

Fig. 8 Truthfulness of $IMC^2$

vary the value of the key parameters to explore the impacts of these parameters.

We first measure the performance of *DATE* with different number of workers ($n$), number of tasks ($m$), priori probability of dependence ($\alpha$), initial accuracy ($\varepsilon$), copy probability ($r$). Then we measure the performance of *Reverse Auction* with different number of workers ($n$), number of tasks ($m$). We set the maximum iterations $\varphi = 100$, and stop the loop if the truth is not changed or the loop reaches the maximum iterations. All the simulations were run on a Centos 7 machine with Intel(R) Xeon(R) CPU E5-2630 2.6GHz and 128 GB memory. Each measurement is averaged over 100 instances.

*B. Evaluation of DATE*

First of all, we attempt to find the best setting of $\varepsilon$, $\alpha$, and $r$ for *DATE*. We fix $r = 0.2$ at a default value, and vary both $\varepsilon$ and $\alpha$ from 0.1 to 0.9. Fig.3 shows that the *precision* fluctuates slightly between 0.82 and 0.92. Thus, *DATE* is not sensitive to initial setting of both $\varepsilon$ and $\alpha$. In our simulations, we set $\alpha = 0.2$ and $\varepsilon = 0.5$ since this setting can obtain the highest *precision* of 0.92. However, as shown in Fig. 3, the *precision* increases significantly when we increase $r$ from 0.1 to 0.4. The *precision* and becomes convergence when $r$ is more than 0.4. The setting of $r$ may be influenced largely by the data set adopted, especially, the number of copiers. We set $r = 0.4$ in our simulations.

Fig.4 compares the *precision* achieved by the *DATE* against the benchmark algorithms. *DATE* can calculate the workers' dependencies, thereby obtaining higher *precisions* (more than 0.85 in all cases) than those of *MV* and *NC* (with average improvement 8.4% and 7.4%, respectively). *ED* outperforms the *DATE* in terms of *precision* (with average improvement 0.8%) since it enumerates all possible dependence for each worker with others when calculating the probability of providing each possible value independently. However, as we shown later, *ED* takes much more running time than *DATE*. Based on the results of Fig.4(a), the *precision* decreases when the task increases. In our simulations, we select the tasks based on the index in the increasing order from the data set. In the data set adopted, the tasks with small index are performed by more workers. This means that fewer values can be used to estimate the truth for the later tasks. Therefore, the *precision* decreases slightly when the number of tasks increases. From Fig.4(b), we can see that the all algorithms obtains the higher *precisions* when the worker increases. This is because the algorithms can estimate the truth from more responses for the tasks.

Fig.5 depicts the running time of all algorithms. It can be seen that the running time of all algorithms increase with the increase both of tasks and workers. Intuitively, the running time of *ED* increases faster than other algorithms since *ED* calculates all possible dependencies of workers, which leads to the complexity of exponential time. For the setting $n$=120, $m$=300, our *DATE* only takes 42.6% of running time comparing with *ED*.

*C. Evaluation of Reverse Auction*

Fig. 6 depicts the social cost of *Reverse Auction*, *GA* and *GB* with different number of tasks and workers. The social cost increases with increasing tasks since there will be more workers to be selected as winners in order to complete the tasks. On the contrary, the social cost decreases with increasing workers. This is because we can find more workers with high-accuracy and lower bid price to perform the same task. The *Reverse Auction* can obtain the lowest social cost comparing with *GA* and *GB* (with average decrease 59.4% and 40.2%, respectively) since *Reverse Auction* can output the social cost with guaranteed approximation.

From Fig.7, we can see that the running time of *Reverse Auction*, *GA*, *GB* increase with the increase both of tasks and workers. This is consistent with our time analysis in Lemma 1. It is not difficult to obtain the time complexity $O(n^3)$ of *GA* and $O(n^2)$ of *GB*, respectively, of which both are lower than $O(n^3m)$ of *Reverse Auction*. Thus the running time of *GA* and *GB* is lower than *Reverse Auction*.

We verified the truthfulness of $IMC^2$ by randomly picking two workers (ID=26 and ID=58) and allowing them to bid prices that are different from their true costs. We illustrate the results in Fig.8. We can see that the worker 26 always obtain its maximum utility of 5 if bidding its real cost $c_{26}=3$. Accordingly, the loser 58 always obtains nonnegative utility if he/she bids truthfully ($b_{58}=c_{58}=8$).

## VIII. RELATED WORK

### A. Truth Discovery in Crowdsourcing

For the paradigm of crowdsourcing, a large body of work on truth discovery has been proposed in the literature. In [31], Miao *et al.* propose the first privacy-preserving truth discovery framework called *PPTD*. *PPTD* relies on the threshold homomorphic cryptosystem to protect the confidentiality of workers' values and weights. Tang *et al.* [32] propose non-interactive privacy-preserving truth discovery which protect workers' data while enabling truth distillation. Xiao *et al.* [37] propose *BUR* protocol which can recruit nearly the minimum number of workers while ensuring that the total accuracy of each task is no less than a given threshold. Wu *et al.* [36] design an unsupervised learning approach to quantify the workers' data qualities and long-term reputations, and exploit an outlier detection technique to filter out anomalous data items. However, all of these studies do not consider the incentive to the workers.

Jin *et al.* [35] propose an integrated framework for multi-requester mobile crowdsourcing systems, called *CENTURION*, consisting of a truth discovery mechanism and an incentive mechanism. The truth discovery mechanism takes workers' reliability into consideration, and calculates highly accurate aggregated results for the requesters. However, they don't consider the crowdsourcing systems with copiers.

### B. Quality-aware Incentive Mechanims in Crowdsourcing

Aware of the importance of stimulating worker participation, various quality-aware incentive mechanisms have been proposed for mobile crowdsourcing systems. In [16], Jin *et al.* propose *INCEPTION*, a novel MCS system framework that integrates the incentive, data aggregation, and data perturbation mechanisms. Wang *et al.* study the problem of measuring workers' long-term quality and they propose *MELODY* [17], a long-term dynamic quality-aware incentive mechanism for crowdsourcing. Wen *et al.* propose an incentive mechanism based on a *Quality Driven Auction* [18], where the worker is paid off based on the quality of sensed data instead of working time. Jin *et al.* [19] design an incentive mechanisms based on reverse combinatorial auctions, and incorporate the *Quality of Information* (QoI) of workers into the incentive mechanism. However, all these studies do not consider the dependence of workers.

Overall, there is no off-the-shelf mechanism designed in the literature, which considers both dependence and accuracy of workers.

## IX. CONCLUSION

In this paper, we have designed a two-stage incentive mechanism for truth discovery in crowdsourcing with copiers. In truth discovery stage, we calculate the dependence for each pair of workers based on the Bayesian analysis, and estimate the truth for each task based on both the dependence and accuracy of workers. In reverse auction stage, we develop a greedy algorithm to maximize the social welfare such that all tasks can be completed with the least confidence for truth discovery. Through both rigorous theoretical analysis and extensive simulations, we have demonstrated that the proposed incentive mechanisms achieve computational efficiency, individual rationality, truthfulness, and guaranteed approximation. Moreover, our truth discovery method shows prominent advantage in terms of precision when there are copiers in the crowdsourcing systems.

ACKNOWLEDGMENT

This work has been supported in part by the NSFC (No. 61872193, 61872191, 61872197), and NSF (No. 1444059, 1717315).